\documentstyle[11pt]{article}
\input psfig
\long\def\comment#1{}
\begin{document}
\title{Paradox of Quantum Information}
\author{Subhash Kak\\
Department of Electrical \& Computer Engineering\\
Louisiana State University,
Baton Rouge, LA 70803, USA}
\maketitle

\begin{abstract}
The notion of quantum information related to the two different perspectives
of the global and local states
is examined.
There is circularity in the definition of quantum information because
we can speak only of the information of systems that have been
specifically prepared.
In particular, we examine 
the final state obtained
by applying unitary transformations on a single qubit that belongs
to an
entangled pair. 

\end{abstract}

\thispagestyle{empty}

An unknown quantum state is unknowable. One may postulate
it to be in a pure or mixed state although there is no
way to determine the type. Interactions with
the state lead to its probabilistic collapse.

One must therefore speak only of systems whose states
have been specifically prepared as pure states. However, 
quantum gates to accomplish this are liable to have
their own imprecision. The pure state 
has uncertainty associated with it\cite{Ka99}.

There is circularity in the consideration of quantum
information. Unlike information in the classical sense where
it
may be defined as characteristics of an unknown 
system, obtained through various measurements, we can only
speak of prepared states (in principle, limited by equipment
precision and measurement accuracy) and then ask questions
regarding its components after certain operations have
been applied on it.

Consider entangled particles\cite{Be95}. Given a specific pair
produced by a physical process, there
is no way to characterize it accurately. For example, we
wouldn't know if it is in a pure state or a mixed stated,
or know the nature of mixedness, if it is mixed.
We must begin by preparing particles in the pure
$|00\rangle$ state and then, using an appropriate 2-qubit gate, 
steer it into the entangled pair in a pure state.

One may view the situation of the entangled particles from the two perspectives
of the system and the individual particles.
In the first, the state of the two particles is considered jointly
(Perspective G related to the generator of the entangled pairs).
In the second perspective, one may consider each particle separately,
without the knowledge that it is entangled with another particle.
From this perspective (corresponding to Alice or Bob), consistency demands
that 
operations on their qubits will provide the same results
as when the particles are unentangled and in an appropriate
mixed state.

Consider the entangled pure state
 $ | \psi\rangle =  \frac{1}{\sqrt 2} (| 00 \rangle +  | 11\rangle)$.
Both particles (qubits) have equal probability of
being 0 or 1.
We may assume that the particles have traveled apart to A and B, so that
one may operate on each separately, without influencing the other.
Consider now the unitary transformation

\vspace{0.2in}
 $H = \frac{1}{\sqrt 2}                       \left[ \begin{array}{cc}
1 & ~1  \\
1 & -1  \\
\end{array} \right]$

Someone who doesn't know
of the entanglement of the particles may
consider the particle at A to be a pure state 
 $  \frac{1}{\sqrt 2} (| 0 \rangle +  | 1\rangle)$.
Then the application of
$H$ would rotate it to the state $|0\rangle$. The state
of the two particles jointly is
 $  \frac{1}{\sqrt 2} (| 00 \rangle +  | 01\rangle) =
  |0\rangle ( \frac{1}{\sqrt 2} (| 0 \rangle +  | 1\rangle))$
Applying $H$ on the second particle, one obtains 
 $| 00\rangle$. This is incorrect from the Perspective G and it
is the expectation based on the individual's lack of
knowledge regarding the system.

From the perspective of G, we know that the entangled particles cannot
individually be in pure states.
The individual particles are
mixed, although the entangled state is a pure state.
As a mixture, the transformation $H$ leads to the state for the pair:

\vspace{0.2in}
 $  \frac{1}{2} (| 00 \rangle +  | 01\rangle + | 10\rangle - | 11\rangle) $

\vspace{0.2in}
The $H$ transformations on the particles available to A and B may be
written as:

\vspace{0.2in}
 $H(A) = \frac{1}{\sqrt 2}                       \left[ \begin{array}{cccc}
1 & ~0 & 1 & 0 \\
0 & ~1 & 0 & 1 \\
1 & ~0 & -1 & 0 \\
0 & ~1 & 0 & -1 \\
\end{array} \right]$

\vspace{0.2in}
 $H(B) = \frac{1}{\sqrt 2}                       \left[ \begin{array}{cccc}
1 & ~1 & 0 & 0 \\
1 & -1 & 0 & 0 \\
0 & ~0 & 1 & 1 \\
0 & ~0 & 1 & -1 \\
\end{array} \right]$

Applying $H(A)$ on the entangled pair
 yields the state
 $  \frac{1}{2} (| 00 \rangle +  | 01\rangle + | 10\rangle - | 11\rangle) $
and applying $H(B)$ on it returns the state to
 $  \frac{1}{\sqrt 2} (| 00 \rangle +  | 11\rangle)$.

The application of $H(A)$ is reversed by the application of $H(B)$.
Also, $H(A) |\psi \rangle = H(B) | \psi \rangle$.
The $H$ operator applied to either of the qubits creates the same
change in the state.
This is surprising because of the asymmetry of the operation.
It is one of the paradoxical characteristics of quantum information.

The joint state of the two particles may be represented as:
 $  \frac{1}{2} ((| 0 \rangle +  | 1\rangle) |0\rangle + (
 | 0\rangle - | 1\rangle)) |1\rangle $
 $=  \frac{1}{2} ( |0\rangle (| 0 \rangle +  | 1\rangle)  + |1\rangle (
 | 0\rangle - | 1\rangle) )  $.
One may view the situation to be either as the particle
at A to be in the mixed state $ \frac{1}{\sqrt 2} |0\rangle + |1\rangle$ and  $ \frac{1}{\sqrt 2}|0\rangle - |1\rangle$
 with the particle
at B to be in the mixed state $|0\rangle$  and $|1\rangle$ or the other way round, equivalently,
or as a combination of the two.

\paragraph{Mode change}

From an information point of view, the nature of the state of
the qubit, whether
mixed or pure, is significant.
When appropriate gate transformations are applied, the states
of the individual qubits would transition from mixed to
pure\cite{Ka03}.
This would happen when the gate removes the
entanglement between the particles.
Such a transition may be measurable under certain conditions
and be used to communicate information
without collapsing the state function.

For example, if the output is mixed as in the case of
the entangled particles, $H$ transformation on it
will leave its state unchanged and a detector after the $H$ operator
will detect $0$ and $1$ with equal probability.
On the other hand, $H$ transformation on the pure state
(created after the entanglement has been removed)
$\frac{1}{\sqrt 2} (| 0 \rangle +  | 1\rangle)$
will change it to $|0\rangle$.
The detector now will measure $0$ in each case.
Given several copies of such states, one can use the
$H$ transformation to distinguish the mixed and the pure states
and, therefore, detect if the gate operation was applied.

But a mode change cannot take place using a local transformation.
Therefore such information transmission can be successfully
accomplished only across multi-qubit gates.


\end{document}